\documentclass[a4paper,11pt,sans]{article}

\usepackage[top=2cm,bottom=2cm,left=2cm,right=2cm]{geometry}
\usepackage[utf8]{inputenc}
\usepackage{graphicx}
\usepackage{bm}
\usepackage{amsmath}

\title{Increasing dynamic range of NESs by using geometric nonlinear damping}
\author{Etienne Gourc, Pierre-Olivier Mattei, Renaud Cote, Matteo Capaldo}
\date{}

\begin{document}

\maketitle

\abstract{The paper deals with the passive control of resonant systems using nonlinear energy sink (NES). The objective is to highlight the benefits of adding nonlinear geometrical damping in addition to the cubic stiffness nonlinearity. The behaviour of the system is investigated theoretically by using the mixed harmonic balance multiple scales method. Based on the obtained slow flow equations, a design procedure that maximizes the dynamic range of the NES is presented. Singularity theory is used to express conditions for the birth of detached resonance cure independently of the forcing frequency. It is shown that the presence of a detached resonance curve is not necessarily detrimental to the performance of the NES. Moreover, the detached resonance curve can be completely suppressed by adding nonlinear damping. The results of the design procedure are then compared to numerical simulations.}

\section{Introduction}

Nonlinear energy sinks (NES) are strongly nonlinear oscillators that are used to mitigate vibration of a host system through targeted energy transfer. One of the main features of NES, which is a consequence of their essential nonlinear nature is their ability to enter into resonance capture with the system to which they are attached, giving them a broadband capability \cite{vakakis2001energy}. When the primary system is subjected to harmonic excitation, the system exhibits relaxation oscillation referred to as strongly modulated response (SMR) which is also a typical feature of systems with NES \cite{starosvetsky2008strongly}. 
One of the main drawbacks of NES when used to control an harmonically excited system is the possible presence of a high amplitude detached resonance curve. Specific design procedures have been developed to ensure a safe operating region of the system, at the cost of limited performance of the NES \cite{starosvetsky2008response,gourc2014experimental}. 
A promising way to overcome this difficulty is the introduction of nonlinear damping in addition to linear viscous damping. This was first proposed in \cite{starosvetsky2009vibration} where the authors considered piecewise quadratic damping. It was shown that imposing a fairly low damping at low amplitude and a higher damping at high amplitude can lead to a complete elimination of the detached resonance curve. In \cite{andersen2012dynamic}, the dynamics of a NES with geometric nonlinear damping is investigated under transient loading. They showed that the nonlinear damping can induce transient instability caused by a bifurcation to $1:3$ resonance capture. The behaviour of a NES with global and local potential as well as nonlinear damping has been analyzed theoretically in \cite{liu2020dynamic}. Recently, the behaviour of a NES with geometric nonlinear damping used to control a toy model of a floating offshore wind turbine was investigated in \cite{mattei2023}. They confirmed both theoretically and by using extensive numerical simulation that the addition of nonlinear damping can, under certain conditions completely suppress the detached resonance curve. Experimental investigation of nonlinear damping, not in the context of NES, is reported in \cite{mojahed2018strong}, where a nonlinear damping element is realized geometrically at a beam tip.

The objective of the present paper is to investigate the benefits induced by the addition of geometrical nonlinear damping. Section \ref{sec:2} is devoted to the presentation of the studied system. Section \ref{sec:3} is devoted to the theoretical analysis of the system. First the equations of motion are analyzed by using the multiple scales-harmonic balance method (MSHBM) \cite{luongo2012dynamic} allowing us to obtain the expression of the slow invariant manifold (SIM) and the slow flow modulation equation. In the same section, the folded singularities are expressed and interpreted as a grazing flow and conditions for the existence of detached resonance curves are expressed using the singularity theory \cite{cirillo2017analysis,habib2018isolated}. The tuning procedure that allows to distinguish problematic from nonproblematic detached resonance curves is presented in section \ref{sec:4}. 

\section{Description of the system}\label{sec:2}

The system considered in the present paper is the same as in \cite{andersen2012dynamic} and consists of a linear oscillator coupled to an embedded NES with geometric nonlinear stiffness and damping. The equations of motion are given by

\begin{equation}\label{eq:1}
\begin{array}{l}
M\ddot{x}+C\dot{x}+Kx+c\dot{w}+rw^3+dw^2\dot{w}=A\cos(\Omega t)\\
m(\ddot{x}-\ddot{w})-c\dot{w}-rw^3-dw^2\dot{w}=0
\end{array}
\end{equation}

where $x(t)$ and $w(t)$ are the displacement of the linear oscillator and the relative displacement of the NES, respectively. The dots denote differentiation with respect to time $t$. $M$, $C$, and $K$ are the mass, damping and stiffness of the linear oscillator, respectively. $c$, $r$ and $d$ are the damping, cubic nonlinear stiffness and nonlinear damping of the NES, respectively. Nondimensional parameters are introduced as follows

\begin{equation}\label{eq:2}
\begin{array}{c}
\tilde{t}=\omega_1 t, \quad \zeta=\frac{C}{2M\omega_1},\quad \mu=\frac{c}{m\omega_1},\quad \kappa=\frac{r}{m\omega_1^2}\\
G=\frac{A}{M\omega_1^2},\quad \omega=\frac{\Omega}{\omega_1},\quad \lambda=\frac{d}{m\omega_1}
\end{array}
\end{equation} 

where $\omega_1=\sqrt{K/M}$. Substituting into Eq. (\ref{eq:1}) yields to the following nondimensional equations of motion

\begin{equation}\label{eq:3}
\begin{array}{l}
\ddot{x}+2\zeta\dot{x}+x+\epsilon(\mu\dot{w}+\kappa w^3+\lambda w^2\dot{w})=G\cos(\omega t)\\
\epsilon(\ddot{x}-\ddot{w}-\mu\dot{w}-\kappa w^3-\lambda w^2\dot{w})=0
\end{array}
\end{equation}

where the tilde has been dropped for brevity and the dots now represent the derivative with respect to adimensional time. $\epsilon=m/M\ll 1$ is the mass ratio.

\section{Nonlinear analysis}\label{sec:3}

The behaviour of the system is analyzed theoretically using the mixed multiple scales-harmonic balance method (MSHBM) \cite{luongo2012dynamic}. Independent time scales $t_i=\epsilon^i t$ are introduced and the displacements are expanded as

\begin{equation}\label{eq:4}
\begin{array}{l}
x(t;\epsilon)=x_0(t_0,t_1)+\epsilon x_1(t_0,t_1)+\ldots\\
w(t;\epsilon)=w_0(t_0,t_1)+\epsilon w_1(t_0,t_1)+\ldots\\
\end{array}
\end{equation}

The damping of the primary system $\zeta$ as well as the forcing amplitude $G$ are assumed to be small and are scaled such that $\zeta=\epsilon\zeta$ and $G=\epsilon G$. Substituting Eq. (\ref{eq:4}) into Eq. (\ref{eq:3}) and collecting terms of the same power of $\epsilon$ gives

\begin{equation}\label{eq:5}
\begin{array}{ll}
\mathcal{O}(\epsilon^0) : & d_0^2x_0+x_0=0
\end{array}
\end{equation}

\begin{equation}\label{eq:6}
\begin{array}{ll}
\mathcal{O}(\epsilon^1) : & d_0^2x_1+x_1=-2d_0d_1x_0-2\zeta d_0x_0-\mu d_0w_0-\kappa w_0^3-\lambda w_0^2 d_0w_0+G\cos(\omega t_0)\\
& d_0^2(w_0-x_0)+\mu d_0w_0+\kappa w_0^3+\lambda w_0^2d_0w_0=0
\end{array}
\end{equation}

where $d_i^j=\partial^j/\partial t_i^j$. The solution at order $\epsilon^0$ is expressed by

\begin{equation}\label{eq:7}
x_0(t_0,t_1)=\frac{1}{2}X(t_1)\mathrm{e}^{it_0}+\mathrm{c.c.}
\end{equation}

Now we deal with the equations at order $\epsilon^1$. Since the second equation does not admit a closed-form solution, according to the MSHBM, we seek an approximate solution of the form

\begin{equation}\label{eq:8}
w_0(t_0,t_1)=\frac{1}{2}W(t_1)\mathrm{e}^{it_0}+\mathrm{c.c.}
\end{equation}

Substituting Eq. (\ref{eq:7}, \ref{eq:8}) into the second equation of Eq.(\ref{eq:6}) and balancing the terms corresponding to the first harmonic yields

\begin{equation}\label{eq:9}
X=(1-i\mu)W-\frac{1}{4}(3\kappa+i\lambda)W^2W^*
\end{equation}

Equation (\ref{eq:9}) represents the slow invariant manifold (SIM) of the problem.
The behaviour of the system is analyzed in the vicinity of the resonance of the primary system. Accordingly, a detuning parameter is introduced as $\omega=1+\epsilon\sigma$. Substituting Eq. (\ref{eq:7}, \ref{eq:8}) into the first equation of Eq.(\ref{eq:6}) and eliminating secular terms reads

\begin{equation}\label{eq:10}
d_1X=-\zeta X-\frac{1}{2}\mu W+\frac{1}{8}(3i\kappa-\lambda)W^2W^*-\frac{1}{2}iG\mathrm{e}^{i\sigma t_1}
\end{equation}

Going back to true time $t$ and reabsorbing $\epsilon$ the modulation equation is obtained as

\begin{equation}\label{eq:11}
\dot{X}=-\zeta X-\epsilon\left(\frac{1}{2}\mu W+\frac{1}{8}(3i\kappa-\lambda)W^2W^*\right)-\frac{1}{2}iG\mathrm{e}^{i\sigma t}
\end{equation}

Since we are interested in the dynamics of the system under $1:1$ resonance capture, i.e. on the SIM, Eq. (\ref{eq:9}) is substituted into Eq. (\ref{eq:11}). Expressing $W(t)=b(t)\mathrm{e}^{i\beta(t)}$ and splitting into real and imaginary parts gives

\begin{equation}\label{eq:12}
\begin{array}{l}
\dot{b}=\frac{f_1(b,\theta)}{g(b)}\\
\dot{\theta}=\frac{f_2(b,\theta,\sigma)}{g(b)}
\end{array}
\end{equation}

where the new phase variable $\theta(t)=\epsilon\sigma t-\beta$ has been introduced to make the system autonomous. The expressions of $f_1$ and $f_2$ are given in appendix. The fixed points of the system are computed by setting $f_1=f_2=0$.

\subsection{Analysis of the SIM}

Substituting the polar expression of $W(t)$ and $X(t)=a(t)\mathrm{e}^{i\alpha(t)}$ into Eq. (\ref{eq:9}), the real-valued expression of the SIM is obtained as

\begin{equation}\label{eq:13}
A=\frac{1}{16}\left[(3\kappa B-4)^2B+(\lambda B+4\mu)^2B\right]
\end{equation}

where $A=a^2$ and $B=b^2$. The values of local extremums $B_1$ and $B_2$ are found by equating the derivative of the right-hand side of Eq. (\ref{eq:13}) with respect to $B$ to zero, giving

\begin{equation}\label{eq:14}
B_i=\frac{4}{3}\frac{6\kappa-2\mu\lambda\mp\sqrt{h}}{9\kappa^2+\lambda^2}
\end{equation}

with

\begin{equation}\label{eq:15}
h=-27\kappa^2\mu^2+\lambda^2\mu^2-24\kappa\lambda\mu+9\kappa^2-3\lambda^2
\end{equation}

\begin{figure}
\begin{center}
\includegraphics[scale=1]{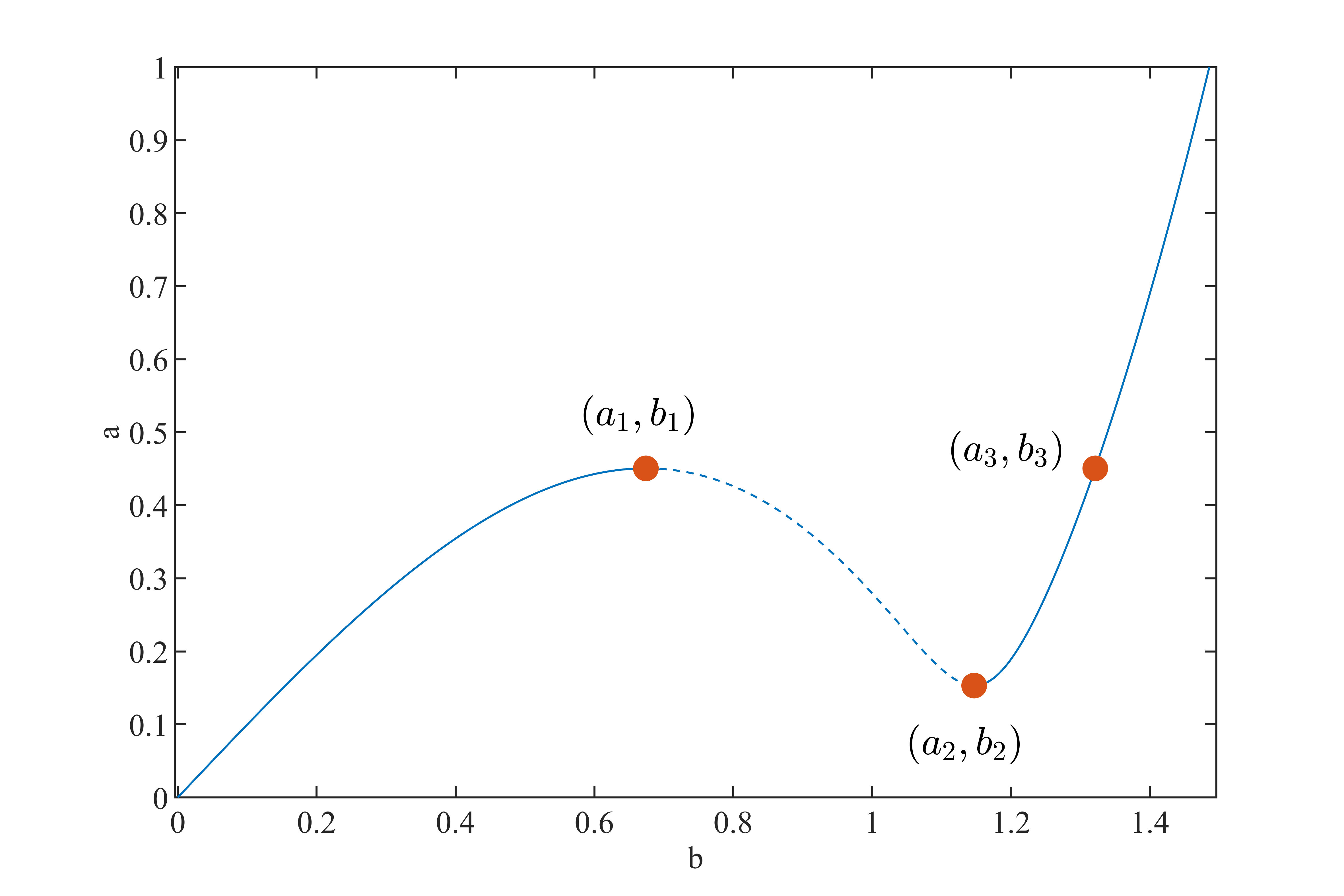}
\caption{SIM for $\mu=0.1$, $\kappa=1$ and $\lambda=0.1$. $a$ and $b$ are the amplitudes of oscillation of the primary system and of the relative displacement of the NES, respectively. Solid and dashed lines correspond to stable and unstable branches, respectively.}
\label{fig:1}
\end{center}
\end{figure}

Note that the SIM admits extremums only if $h>0$. It is well known that this topology of SIM can give rise to strongly modulated response. An example of SIM for $\mu=0.1$, $\kappa=1$ and $\lambda=0.1$ is depicted in Fig. \ref{fig:1}. 

\subsection{Folded singularities}

In addition to the classical fixed points of the slow flow system, computed by setting $f_1=f_2=0$ and $g\neq 0$, it is well known that systems with NES also admit singular fixed points denoted as folded singularities \cite{starosvetsky2008strongly}. The folded singularities are found by setting $f_1=f_2=g=0$. Physically, folded singularities allow the flow to jump from the lower to the upper branch of the SIM.
On the phase plane $(\theta,b)$, the transition from a situation where the slow flow remains on the lower part of the SIM to a situation where the flow can jump to the higher branch of the SIM corresponds to a grazing flow at $b=b_1$, that is

\begin{equation}\label{eq:16}
\left.\frac{db}{d\theta}\right|_{b=b_1}=0
\end{equation}

with

\begin{equation}\label{eq:17}
\frac{db}{d\theta}=\frac{\frac{db}{dt}}{\frac{d\theta}{dt}}=\frac{f_1}{f_2}
\end{equation}

\begin{figure}
\begin{center}
\includegraphics[scale=1]{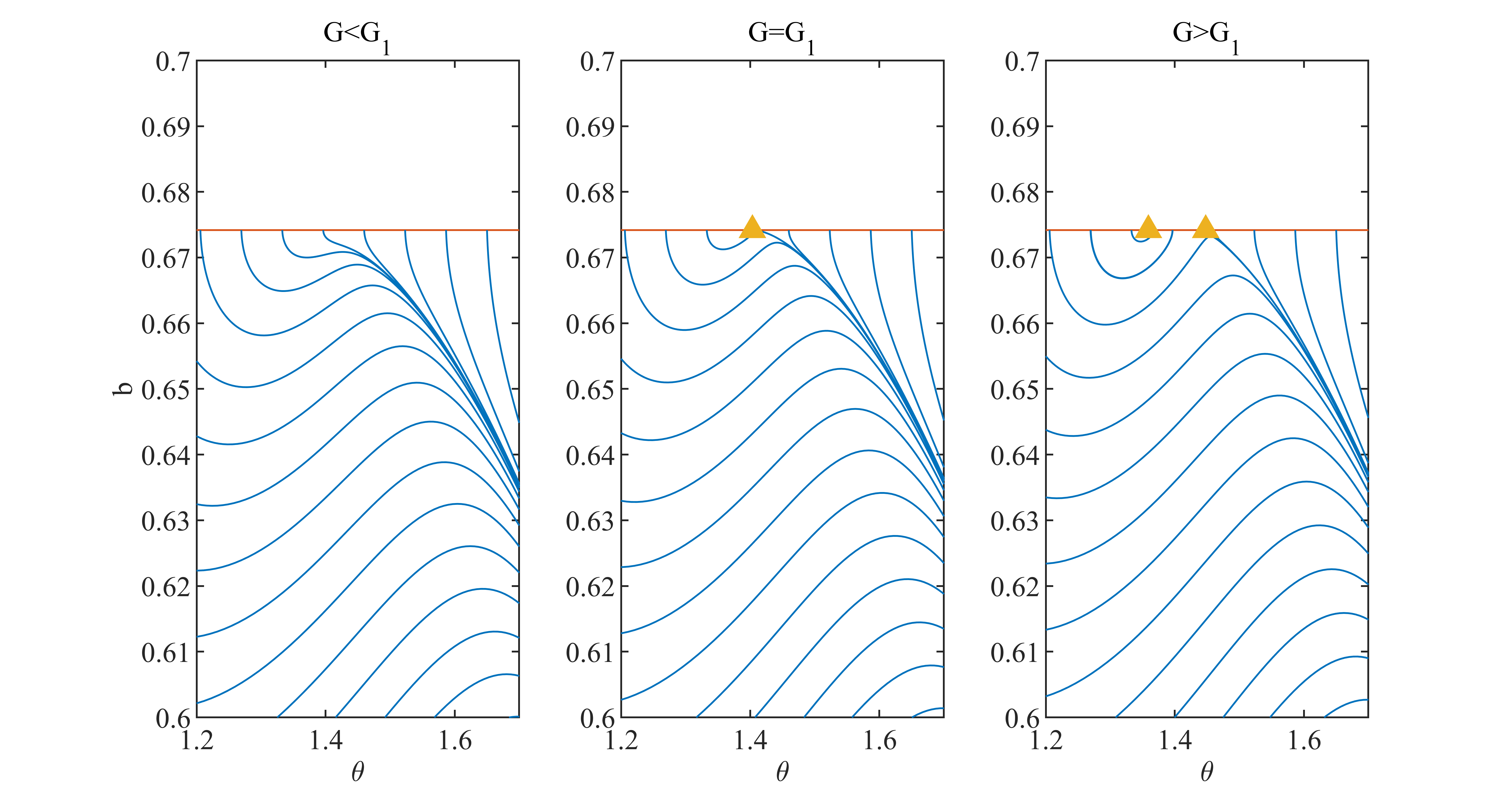}
\caption{Phase plot for $\mu=0.1$, $\epsilon=0.01$, $\zeta=0.01$, $\kappa=1$, $\lambda=0.1$ and $\sigma=1$ for three values of $G$ around $G_1=0.0101$. The solid red line corresponds to $b=b_1$ and orange triangles to the folded singularities.}
\label{fig:2}
\end{center}
\end{figure}

Such that the grazing condition simply reads $f_1(b_1,\theta)=0$. From this relation, it is possible to express the minimal amplitude $G_{\mathrm{fs}_1}$ that leads to the existence of folded singularities. Note that since $f_1$ is independent of $\sigma$, $G_{\mathrm{fs}_1}$ is valid for the whole excitation frequency range. An example of grazing bifurcation giving rise to the birth of a pair of folded singularities is depicted in Fig. \ref{fig:2}.

For completeness, the folded singularities on the upper branch of the SIM at $b=b_2$ can be expressed similarly, i.e. $f_1(b_2,\theta)=0$, and the critical forcing amplitude yielding to their appearance is denoted by $G_{\mathrm{fs}_2}$.

\subsection{Detached resonance curve}

As mentioned earlier, systems with NES can exhibit a detached resonance curve that can yield to high amplitude oscillations and therefore deteriorate the performance of the NES. A polynomial expression for the amplitude of the fixed points $b$ can be obtained by solving $f_1=f_2=0$ for the trigonometric terms and using trigonometric identity. The fixed points $b$ are now found by solving 

\begin{equation}\label{eq:18}
f_3(b,\sigma)=0
\end{equation}

\begin{figure}
\begin{center}
\includegraphics[scale=1]{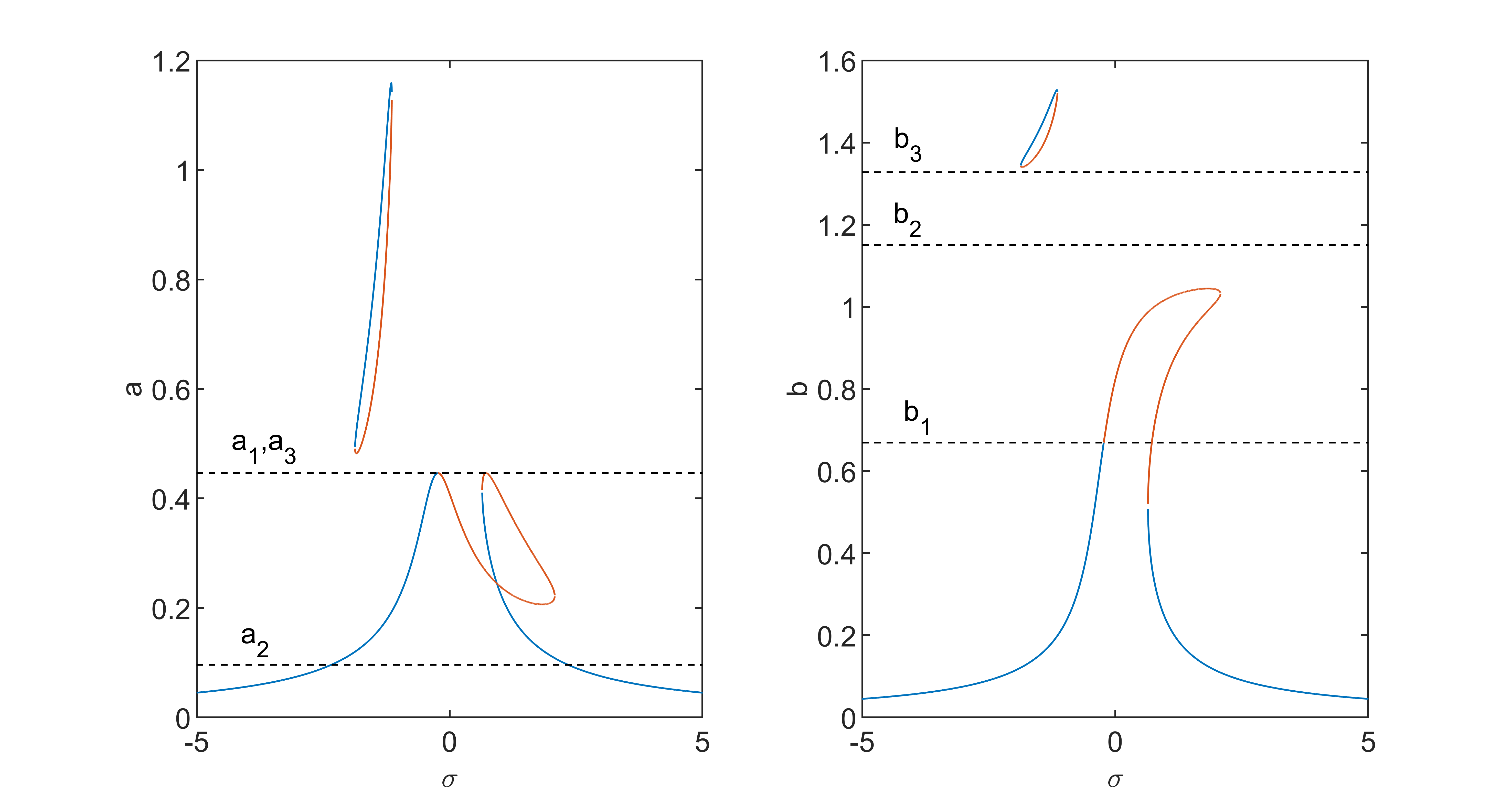}\\
\includegraphics[scale=1]{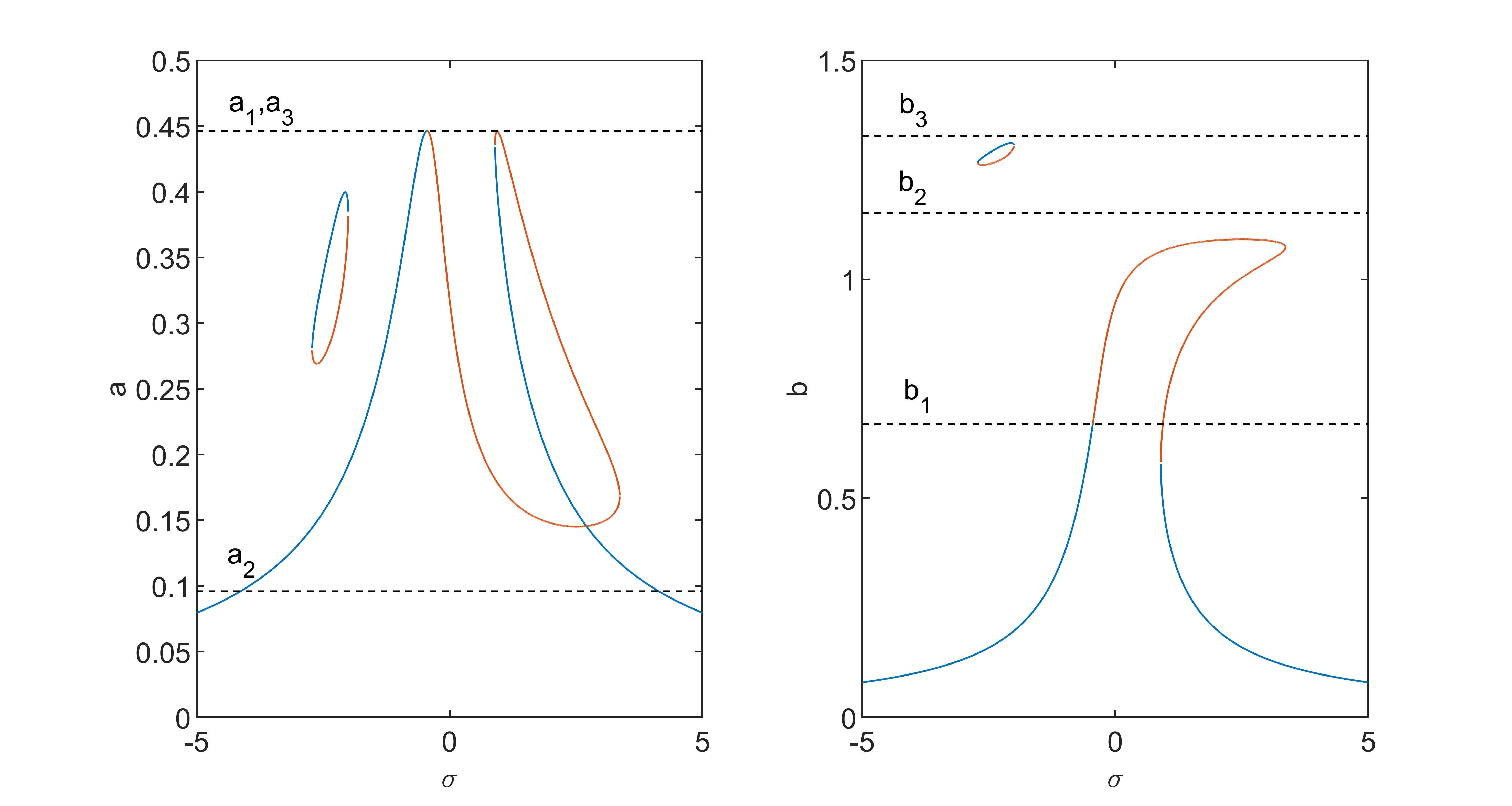}
\caption{Frequency response curve for $\epsilon=0.01$, $\lambda=0.1$, $\kappa=1$, $\mu=0.05$. Up : $\zeta=0.001$, $G=0.0045$, down : $\zeta=0.005$, $G=0.08$. Blue and orange lines correspond to stable and unstable solutions, respectively. Dashed lines indicate the singular points of the SIM.}
\label{fig:3}
\end{center}
\end{figure}

Singularity theory is a powerful tool to classify the different topologies of the frequency response curve \cite{cirillo2017analysis}. Here, we are particularly interested in isola and simple bifurcation singularities, that are responsible for the creation of the isola and the merging of the detached resonance curve with the main frequency response curve, respectively. The defining conditions for these singularities are given by

\begin{equation}\label{eq:19}
\begin{array}{llllll}
\text{isola : }&f_3=0, & \frac{\partial f_3}{\partial b}=0, & \frac{\partial f_3}{\partial\sigma}=0, & \frac{\partial^2f_3}{\partial b^2}\neq 0,& \mathrm{det}(d^2f_3)>0\\
\text{simple bifurcation : }&f_3=0, & \frac{\partial f_3}{\partial b}=0, & \frac{\partial f_3}{\partial\sigma}=0, & \frac{\partial^2f_3}{\partial b^2}\neq 0,& \mathrm{det}(d^2f_3)<0
\end{array}
\end{equation}

where $d^2f_3$ is the Hessian matrix of $f_3(b,\sigma)$. Closed-form solutions are not available, however, Eq. (\ref{eq:19}) can be combined to express a sixth-order polynomial in $b^2$ that can be efficiently solved using any root-finding algorithm and the corresponding forcing amplitude can be retrieved by using the fixed-point equation. Remarkably, the obtained result is independent of $\sigma$. The critical forcing amplitudes corresponding to isola and simple bifurcation singularities are denoted $G_\mathrm{drc}$ and $G_\mathrm{sb}$, respectively. Note however that the presence of a detached resonance curve does not necessarily yields to higher amplitude compared to the principal resonance curve. An example is depicted in Fig. \ref{fig:3} for two different sets of parameters. 

Clearly, in the upper case, the detached resonance curve is problematic as it yield to high amplitude oscillations. Problematic cases can be easily detected by looking at the amplitude $b$ at which the detached resonance curve is created. Exploiting the topology of the SIM, if the isola singularity is located at $b>b_3$ it will cause high amplitude oscillations of the primary system, as shown in Fig. \ref{fig:3}.

\section{NES sizing}\label{sec:4}

In this section, the sizing of the NES is addressed and can be formulated as follows:
\emph{Given a maximum allowable amplitude of the primary system, what are the parameters of the NES that maximize the forcing amplitude?}

\begin{figure}
\begin{center}
\includegraphics[scale=1]{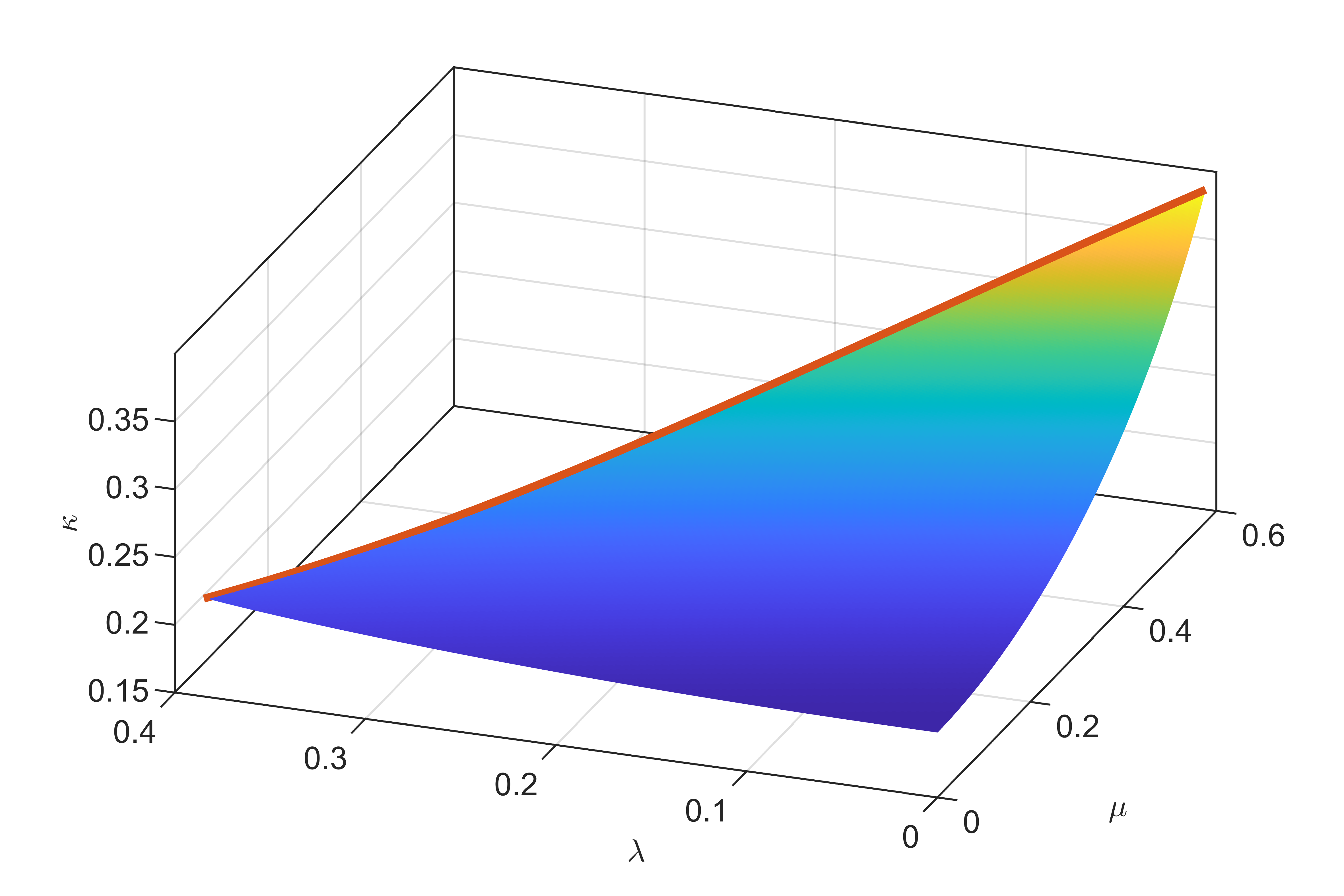}
\caption{Evolution of $\kappa$ as a function of $\mu$ and $\lambda$ for $a_1=1$. Red line corresponds to the limit where the SIM admit a double root.}
\label{fig:4}
\end{center}
\end{figure}

It is known that the the control mechanism of NES under harmonic forcing involves strongly modulated response (SMR) regime. This regime corresponds to relaxation oscillation where the dynamics successively jumps between the two stable branches of the SIM. If no other attractors (i.e. fixed points) are present, the maximum amplitude of the primary system is directly governed by the shape of the SIM and corresponds to the point labelled $a_1$ in Fig. \ref{fig:1}.
The first step of the design procedure consists in determining the nonlinear stiffness coefficient $\kappa$ to fix the maximum amplitude $a_1$. This is done by using the expression of the SIM Eq. (\ref{eq:13}) at $b=b_1$. Notice that the obtained expression is invariant with respect to $a_1$ if $\kappa=\tilde{\kappa}/a_1^2$ and $\lambda=\tilde{\lambda}/a_1^2$ such that without loss of generality we choose $a_1=1$.
The value of $\kappa$ as a function of the linear and nonlinear damping $\mu$ and $\lambda$ allows to reduce the design space by one and is depicted in Fig. \ref{fig:4}.

The second step consists in identifying the parameter space where $a\leq a_1$. Exploiting the shape of the SIM, an equivalent condition in terms of the relative amplitude of the NES reads $b\leq b_3$. 
A necessary but not sufficient condition to express the maximum forcing amplitude $G_{\mathrm{max}}$ such that $b\leq b_3$ is given by

\begin{equation}\label{eq:20}
\left.\frac{\partial b}{\partial\sigma}\right|_{b=b_3}=0
\end{equation}

Using implicit differentiation, the condition in Eq. (\ref{eq:20}) becomes

\begin{equation}\label{eq:21}
f_3=0,\quad \left.\frac{\partial f_3}{\partial\sigma}\right|_{b=b_3}=0
\end{equation}
 
However, as mentioned above, this condition is not sufficient to guarantee that $a\leq a_1$ due to the possible presence of a detached resonance curve at higher amplitude. As explained in the previous section, problematic detached resonance curves are easily identified by looking at the amplitude $b$ corresponding to the isola singularity. If $b>b_3$, the resulting isola will lead to an amplitude of oscillations greater than $a_1$. On the contrary, if a detached resonance curve is created in the lower portion of the SIM ($b<b_3$), while the growth of this detached resonance curve until $a=a_1$ will be captured by Eq. (\ref{eq:21}). 

\begin{figure}
\begin{center}
\includegraphics[scale=1]{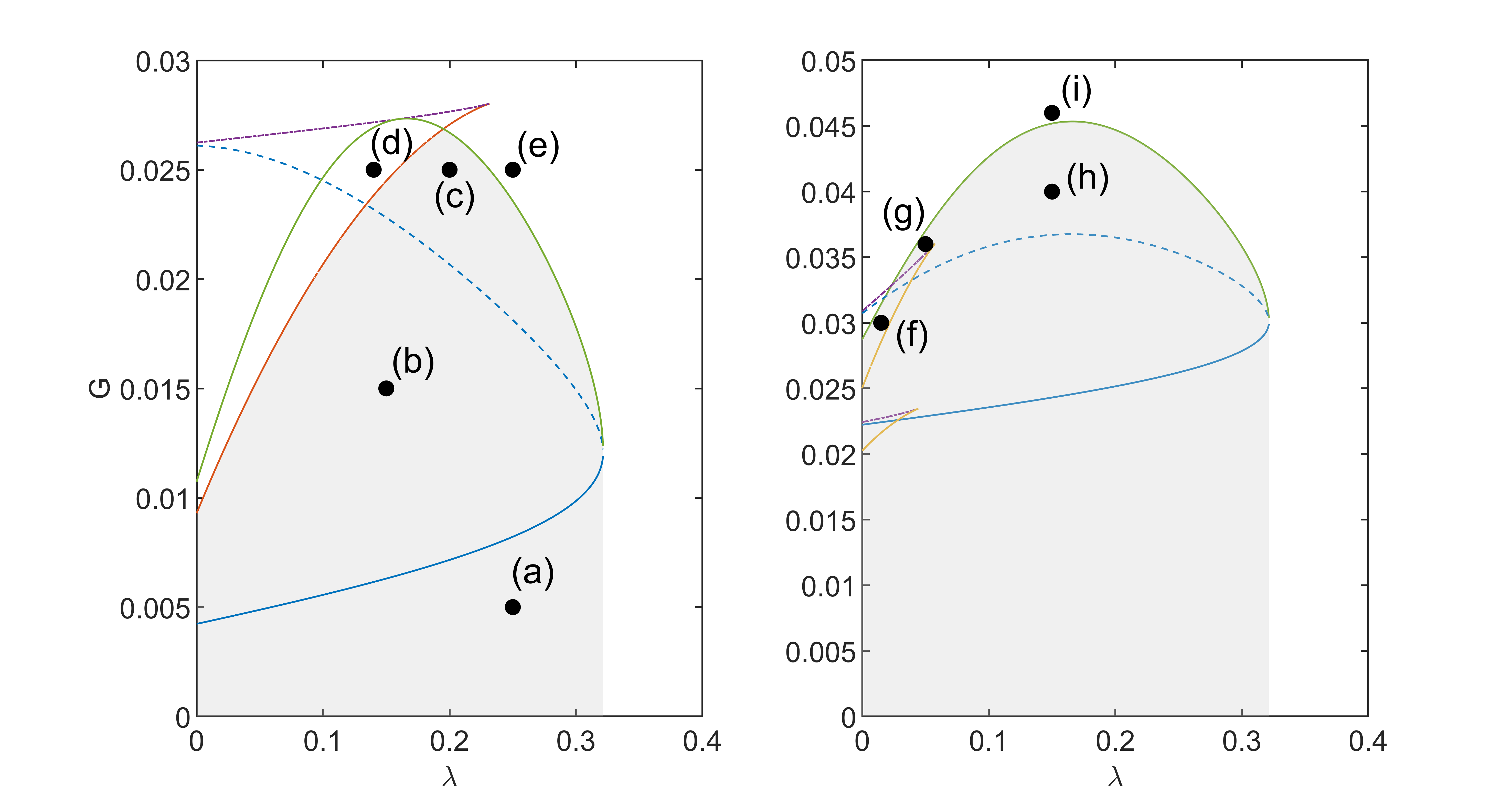}
\caption{Sizing diagrams showing the evolution of the critical forcing amplitudes $G$ as a function of the nonlinear damping coefficient $\lambda$, for $\epsilon=0.01$, $\mu=0.1$. Left: $\zeta=0.001$, right: $\zeta = 0.01$. Solid and dashed blue lines correspond to lower and upper folded singularities, respectively. Red and yellow lines correspond to problematic and nonproblematic isolas, purple lines correspond to simple bifurcation singularities. Green lines correspond to the maximum forcing amplitude $G_\mathrm{max}$.}
\label{fig:5}
\end{center}
\end{figure}

Examples of obtained sizing diagrams are depicted in Fig. \ref{fig:5}. In both cases $\epsilon=0.01$ and $\mu=0.1$. On the left diagram $\zeta=0.001$ while on the right diagram $\zeta=0.01$. Solid and dashed blue lines correspond to folded singularities occurring on $b=b_1$ and $b=b_2$, respectively. Solid green lines correspond to the maximum forcing amplitude $G_\mathrm{max}$. Orange and yellow lines correspond to problematic and non-problematic isola singularities, respectively and finally purple dash-dotted lines correspond to simple bifurcation singularity that yields to the merging of the detached resonance curve with the main resonance curve. The safe region of operation of the NES is indicated by the grey area.

\begin{figure}
\begin{center}
\includegraphics[scale=1]{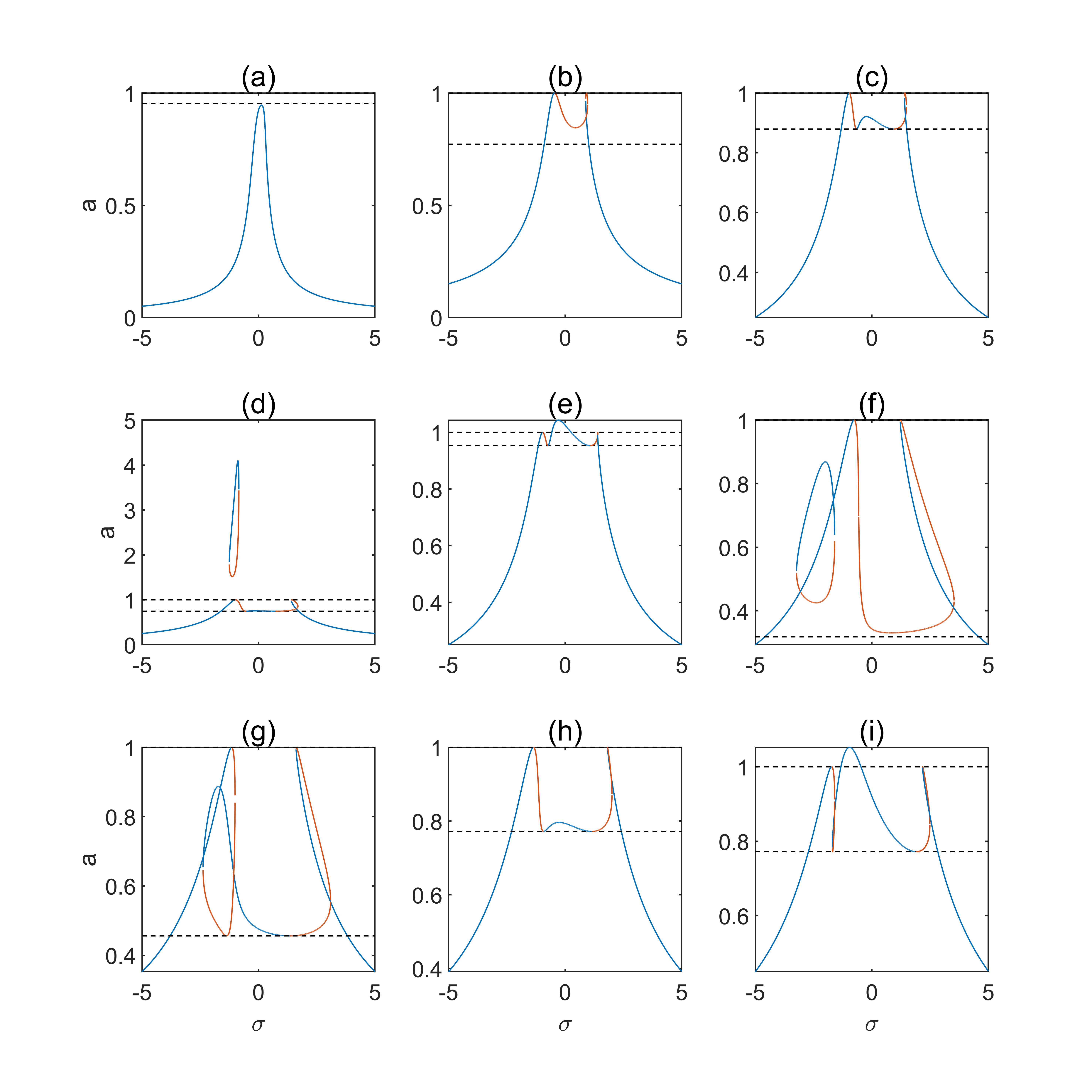}
\caption{Frequency response curve of the primary oscillator corresponding to points in Fig. \ref{fig:5}.}
\label{fig:6}
\end{center}
\end{figure}

Frequency response curves corresponding to different areas of the sizing diagram are shown in Fig. \ref{fig:6}. Below the first folded singularity (Fig. \ref{fig:6}(a)), the SMR regime is not possible and the NES behaves quasi-linearly. In the zone corresponding to Fig. \ref{fig:6}(b), i.e. between the two folded singularities and below any detached resonance curve, the periodic solutions around $\sigma=0$ are unstable and the only possible response is SMR. When the forcing amplitude is increased above the second folded singularity and below the maximum forcing amplitude (Fig. \ref{fig:6}(c,h)), the SMR regime can be replaced by stable periodic response, but still satisfying the design criteria (i.e. $a<1$). Fig. \ref{fig:6}(d,f) again illustrates problematic and nonproblematic detached resonance curves, similar to Fig. \ref{fig:3}. In the case depicted in Fig. \ref{fig:6}(f), the presence of the detached resonance curve does not restrict the design area. In the case presented in Fig. \ref{fig:6}(g), the forcing amplitude is increased above the simple bifurcation singularity corresponding to the merging of the detached resonance curve. In this case, the design criteria is still satisfied. In the case depicted in Fig. \ref{fig:6}(e,i), the parameters of the system are above the maximum forcing amplitude $G_{\mathrm{max}}$, and even if no detached resonance curves are present, the stable periodic solution exceeds the maximum amplitude (i.e. $a>1$). 
Nevertheless, for both sizing diagrams, it can be noticed that the addition of nonlinear damping significantly increases the dynamic range of the NES.

\begin{figure}
\begin{center}
\includegraphics[scale=1]{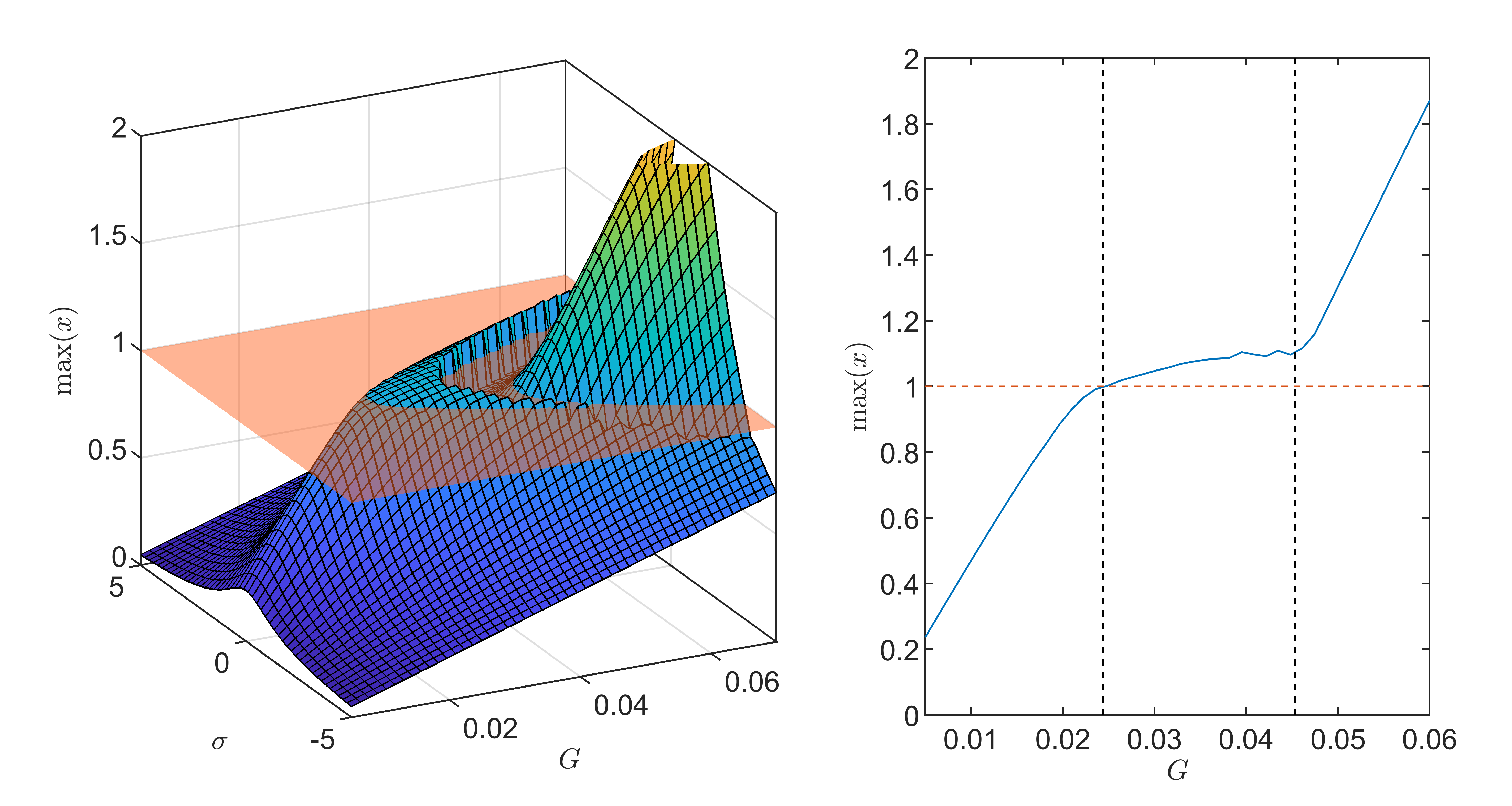}
\caption{Amplitude of the primary oscillator as a function of the frequency detuning $\sigma$ and forcing amplitude $G$ for $\epsilon=0.01$, $\zeta=0.01$, $\mu=0.1$, $\lambda=0.16$, and $\kappa=0.214$. Red plane (dashed line) corresponds to the theoretical maximum allowed amplitude and the dashed vertical lines correspond to $G_{\mathrm{fs}_1}$ (left) and $G_{\mathrm{max}}$ (right).}
\label{fig:7}
\end{center}
\end{figure}

The proposed sizing procedure is compared with numerical simulation in Fig. \ref{fig:7} where the maximum amplitude of the primary oscillator obtained from numerical integration is depicted for various excitation amplitudes and frequencies for $\epsilon=0.01$, $\zeta=0.01$, $\mu=0.1$, $\lambda=0.16$, and $\kappa=0.214$. The red plane and dashed lines correspond to the maximum allowed amplitude. The first vertical dashed line corresponds to the critical forcing yielding to the birth of a pair of folded singularities $G_{\mathrm{fs}_1}=0.0244$ while the second vertical dashed line corresponds to the maximum forcing amplitude $G_\mathrm{max}=0.0453$. The location of the control plateau for $G\in[G_{\mathrm{fs}_1},G_\mathrm{max}]$ is in very good agreement with the numerical simulations while the maximum amplitude is underestimated by the theoretical analysis. 

\begin{figure}
\begin{center}
\includegraphics[scale=1]{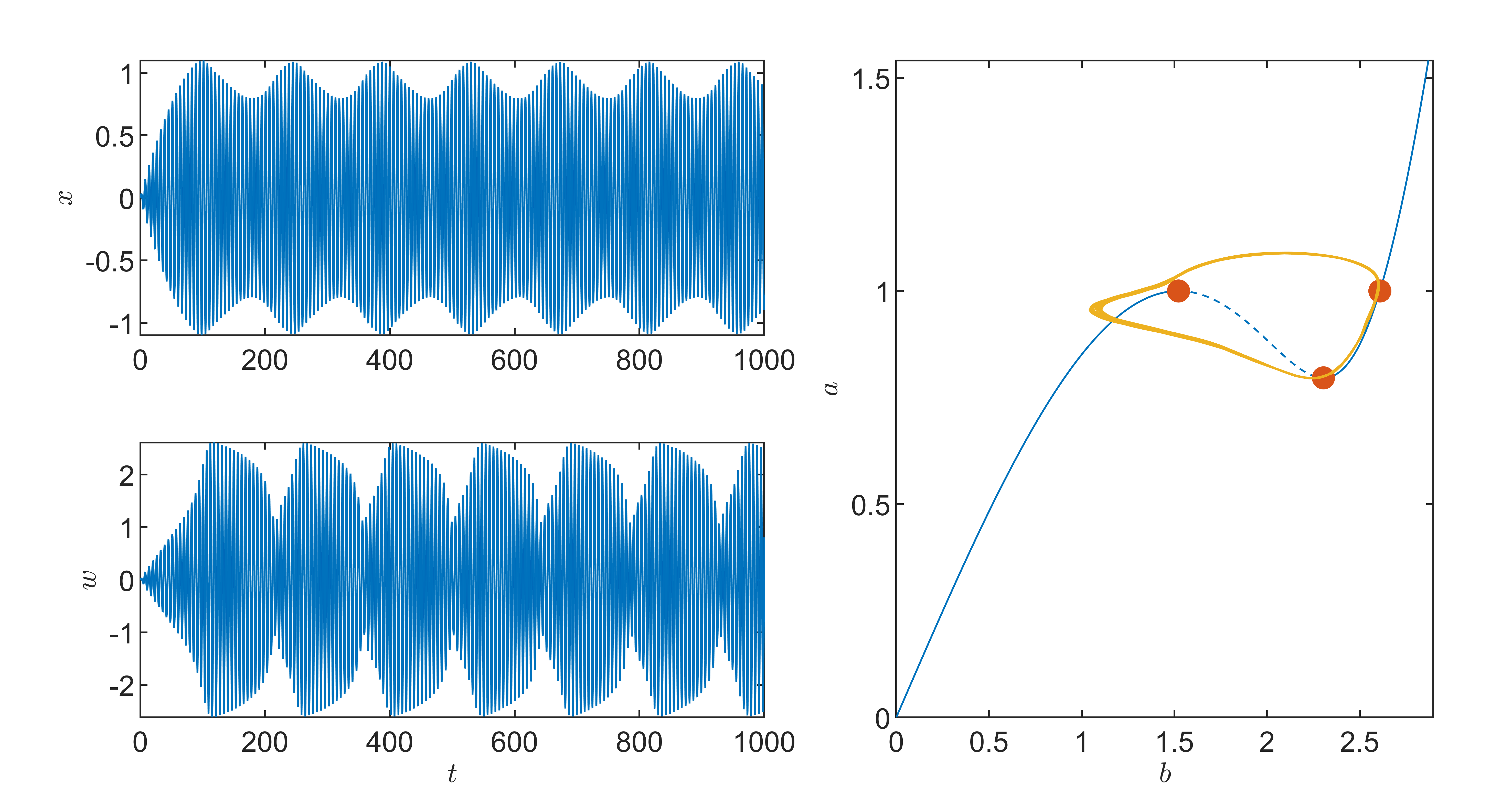}
\caption{Example of strongly modulated response and projection on the SIM obtained for $\epsilon=0.01$, $\zeta=0.01$, $\mu=0.1$, $\lambda=0.16$, $\kappa=0.214$, $\sigma=1.1$ and $G=0.04$.}
\label{fig:8}
\end{center}
\end{figure}

As illustrated in Fig. \ref{fig:8} where $G=0.04$ and $\sigma=1.1$, this is due to the fact that the jump between the branches of the SIM does not occur as a discontinuous phase trajectory when $\epsilon\neq 0$. Note that this fact has been investigated theoretically in \cite{bergeot2021scaling} by interpreting the jump as a dynamic fold bifurcation.

\section{Conclusion}

The objective of the present paper was to investigate the potential benefits of geometric nonlinear damping on NES. To this end, the dynamics of the system has been analyzed by using the mixed multiple scales - harmonic balance method. A simple design procedure is presented, which can be summarized as follow: \emph{given a maximum allowable amplitude of the primary system, what are the parameters of the NES that maximize the forcing range.} 
The main theoretical findings presented throughout the design procedure are twofold:
\begin{itemize}
\item Exploiting the topology of the SIM, in combination with singularity theory, we are able to distinguish parameters that yield to unacceptable or acceptable detached resonance curves.
\item Again, exploiting the topology of the SIM and the expression of the fixed points, we can express a maximum forcing amplitude independent of the forcing frequency that gives rise to a given amplitude of the host oscillator.
\end{itemize}
The combination of these two criteria allows us to define a safe design space of the NES guaranteeing that the vibration amplitude of the primary oscillator is below a certain threshold.
Finally, this design procedure highlights the benefits of nonlinear damping with a significant increase of the maximum allowable forcing amplitude.

\section*{Appendix A}\label{appendixA}

The expressions of $f_1$, $f_2$ and $g$ given in Eq. (\ref{eq:12}) are given by

\begin{equation}
\begin{array}{lll}
f_1(b,\theta)&=&b\left(-\zeta\left(9\kappa^2+\lambda^2\right)b^5+2\left(12\kappa\zeta-\epsilon\lambda-4\lambda\mu\zeta\right)b^3-8\left(2\mu^2\zeta+\epsilon\mu+2\zeta\right)b\right.\\
& & \left.+8G\mu\cos\theta+2G\left(3\kappa b^2+4\right)\sin\theta\right)\\
f_2(b,\theta)&=&3\epsilon(9\kappa^2+\lambda^2)(2\sigma+1)b^5+4\left(\epsilon (8\lambda\mu\sigma-24\kappa\sigma+4\lambda\mu-3\kappa)+4\zeta(\lambda+3\kappa\mu)\right)b^3\\
& & +16\epsilon\left(2\mu^2\sigma+\mu^2+2\sigma\right)b-4G(9\kappa b^2-4)\cos\theta-4G(3\lambda b^2+4\mu)\sin\theta\\
g(b)&=& 2b\left(3(\lambda^2+9\kappa^2)b^4+16(\lambda\mu-3\kappa)b^2+16(\mu^2+1)\right)
\end{array}
\end{equation}

Solving the above equation for $f_1=f_2=0$ yields to

\begin{equation}
\begin{array}{l}
\cos\theta=b\dfrac{3\epsilon\kappa\left(1+2\sigma\right)b^2+2\lambda\zeta b^2+8\left(\mu\zeta-\epsilon\sigma\right)}{4G}\\
\sin\theta=b\dfrac{\epsilon\lambda\left(1+2\sigma\right)b^2-6\kappa\zeta b^2+4\epsilon\mu\left(1+2\sigma\right)+8\zeta}{4G}
\end{array}
\end{equation}

These equation are combined by using trigonometric identity to obtain the fixed point expression, Eq. (\ref{eq:18}) as

\begin{equation}
f_3(b,\sigma)\equiv \cos^2\theta+\sin^2\theta-1
\end{equation}

\bibliographystyle{plain}
\bibliography{bilio}

\end{document}